%% file: HaloFormFactors_111015.tex
\documentclass[
preprint
,showpacs
,showkeys
,nofootinbib
,aps
,amsfonts
,amssymb
,pdftex 
]{revtex4-1}

\usepackage{graphicx} 
\usepackage{amsmath, amssymb, bm} 
\usepackage[dvips]{color}

\usepackage{psfrag,epsfig}
\usepackage{natbib}
\usepackage{hyperref}
\usepackage{slashed}

\def\){\right)} 
\def\({\left(} 
\def\]{\right]} 
\def\[{\left[}

\def\erasepar#1{}

\begin{document}

\title{
Electromagnetic form  factors of one neutron halos with spin $\frac{1}{2}^+$  ground state}

\author{%
Lakma Fernando $^a$}
\email{nfernan1@nccu.edu}

\author{%
Akshay Vaghani $^b$}
\email{av298@msstate.edu}

\author{%
Gautam Rupak $^b$}
\email{grupak@u.washington.edu}

\affiliation{$^a$ North Carolina Central University, 1801 Fayetteville St., Durham, NC 27707, USA}

\affiliation{$^b$ Department of Physics $\&$ Astronomy and 
HPC$^{2}$ Center for Computational Sciences, 
Mississippi State
University, Mississippi State, MS 39762, USA }
\begin{abstract}
The electromagnetic form factors for single neutron halo nuclei $^{11}$Be, $^{15}$C and $^{19}$C are calculated. The calculations are performed in halo effective field theory (EFT) where the halo nuclei are approximated as made of a single neutron and a core.   The form factors depend on the single neutron separation energy, the $s$-wave neutron-core scattering effective range and a two-body current.  The EFT expressions are presented to leading order for  $^{15}$C and next-to-leading order for 
$^{11}$Be and $^{19}$C.

\end{abstract}


\maketitle

\section{Introduction}\label{sec_intro}

In recent years there have been significant progress~\cite{Phillips:2010dt,Hammer:2011ye,Rupak:2011nk,Fernando:2011ts,Lensky:2011he,PhysRevC.86.044608,Acharya:2013nia,Zhang:2014zsa,Ryberg:2014exa,Ryberg:2015lea} in the effective field theory (EFT)  treatment of electromagnetic properties of halo nuclei, following early work in Refs.~\cite{Bertulani:2002sz,Bedaque:2003wa}. Halo nuclei structure and reactions  play an important role in heavy element synthesis in  nuclear astrophysics~\cite{Bertulani:2009mf,Rauscher:2010pu, kawano:1991ApJ372,Wiescher1990,Wiescher1999,Kajino1990}. They provide an unique window into the properties of exotic nuclei near the driplines resembling weakly-bound clusters rather than tightly bound shell-like structures. There is renewed interest in the study of halo nuclei due to the advent of present and planned experiments with high intensity beams of exotic radioactive rare isotopes. Further, the single and two nucleon halo nuclear systems display properties that are universal such as the Efimov effect~\cite{Efimov:1971a,Efimov:1993a,Fedorov:1994,Mazumdar:2000,Yamashita:2005wu,Canham:2008jd}, and can be realized in few-body atomic systems as well~\cite{Kraemer:2006Nat,Braaten:2004a}.

EFT are ideally suited for the study of halo nuclei at low energy.  The clear separation of energy scales -- the small energy required to remove the valance nucleon (or nucleons) and the large energy required to break apart the tightly bound core -- allows one to construct a  low energy EFT. Physical observables are expressed as expansions in the small ratio $Q/\Lambda$ where $Q$ is a momentum associated with the low energy and $\Lambda$ is the momentum associated with the high energy physics. In EFT, the core and loosely  bound particle are considered as  fundamental fields to reduce the complexity of the problem. For example, in $^{15}$C that is represented as a single neutron halo of a $^{14}$C core, the binding momentum $\gamma\sim 46.21$ MeV for valance neutron separation is associated with the soft scale $Q$ whereas the  momentum threshold for pion physics, the excited states of the core $^{14}$C, etc., is identified with the hard scale $
\Lambda\sim 100$-200 MeV~\cite{PhysRevC.86.044608}.  In the EFT, at a given order in the $Q/\Lambda$ expansion all the relevant quantum operators are systematically included.  The theoretical error is estimated from  the higher order terms in the perturbative $Q/\Lambda$ expansion. 
 
In this work we calculate the electromagnetic form factors for $s$-wave spin $\frac{1}{2}$ halo nuclei. The electric form factor for $^{11}$Be 
was studied in Ref.~\cite{Hammer201117}. We include the magnetic form factor as well as apply the analysis to couple of other halo nuclei  $^{15}$C and $^{19}$C.
Form factors of  nuclei have been a longstanding subject of interest in nuclear physics.  Experiments on elastic electron scattering from a nucleus provide essential information about the internal structure of the nucleus, such as charge density and magnetic moment. The form factors are generally written as the ratio of the electron-nuclei cross section to the Mott scattering cross section off point-like particle.  The halo nuclei we consider --  $^{15}$C, $^{11}$Be, and $^{19}$C --  can be analyzed similar to electron-proton scattering.  These nuclei involve  spin $\frac{1}{2}^+$ hadrons as the nuclear target just like the proton. 
The halo nuclei ground states $^{11}$Be, $^{15}$C and $^{19}$C were studied in Refs~\cite{Phillips:2010dt,Hammer:2011ye,PhysRevC.86.044608,Acharya:2013nia}. The construction of the EFT for these nuclei is similar though the power counting that determines the relative sizes of the quantum operators are system specific. The form factor calculation is sensitive to these differences, and they will be discussed when we consider the specific nuclei.

The organization of the paper is as follows. In Section~\ref{sec_theory} we introduce the general formalism for the electric and magnetic form factors. Section~\ref{sec_EFT} introduces the EFT interactions and the form factor calculations. Then we discuss the results for the specific halo systems in Section~\ref{sec_results}.  The power countings in the  halo systems are discussed, and  the corresponding  EFT parameters are chosen.  Conclusions are presented in ~\ref{sec_conclusions}. 

\section{Formalism}\label{sec_theory}

Elastic electron scattering on $\frac{1}{2}^+$ halo  nucleus can be analyzed similar to electron scattering on proton target as both involve spin $\frac{1}{2}$ hadrons.  
The  elastic scattering  amplitude separates into the leptonic and hadronic currents as
\begin{align}
i\mathcal M = [ie \bar\psi_e(-\bm{p}',s')\gamma^\mu\psi_e(-\bm{p},s)]\(-i\frac{g_{\mu\nu}}{q^2}\)
[i\bar \psi_\phi(\bm{p}',a')J_\phi^\nu \psi_\phi(\bm{p},a)],
\end{align}
where $\psi_e(\bm{p},s)$ and $\psi_\phi(\bm{p},s)$ are the electron and halo nucleus Dirac fields with momenta $\bm{p}$ and spins $s$ respectively. The photon momentum $q= p'-p$. Summing over final spins and averaging over initial spins we get
\begin{align}
\frac{1}{2}\frac{1}{2}\sum_{s,s'}\sum_{a,a'}|\mathcal M|^2\equiv \frac{e^2 }{(q^2)^2} g_{\mu\nu} g_{\alpha\beta}L^{\mu\alpha}
T^{\nu\beta},
\end{align}
where the leptonic contribution is written as
\begin{align}
L^{\mu\alpha}= &\frac{1}{2}\sum_{s,s'} [\bar\psi_e(-\bm{p}',s')\gamma^\mu\psi_e(-\bm{p},s)][\bar\psi_e(-\bm{p},s)
\gamma^\alpha\psi_e(-\bm{p}',s')]\nonumber \\
= &\frac{1}{2} \operatorname{Tr}[(-{\slashed p}' +m_e)\gamma^\mu(-\slashed p+m_e)\gamma^\alpha]\nonumber\\
 = & 2[p'^\mu p^\alpha+p'^\alpha p^\mu-(p'\cdot p) g^{\mu\alpha}]+2 m_e^2 g^{\mu\alpha}.
\end{align}
$m_e$ is the electron mass. The hadronic contribution is
\begin{align}
T^{\nu\beta}=\frac{1}{2}\sum_{a,a'}
[\bar\psi_\phi(\bm{p}',a')J^\nu\psi_\phi(\bm{p},a)][\bar\psi_\phi(\bm{p},a)
J^\beta\psi_\phi(\bm{p}',a')]
\end{align}
Current conservation $q_\mu T^{\mu\nu}=0=q_\nu T^{\mu\nu}$ and Lorentz invariance restricts the form of the hadronic current to a generic form
\begin{align}
i \Gamma^\mu= &i \bar \psi_\phi(\bm{p}',a')J_\phi^\mu \psi_\phi(\bm{p},a) =  i e Z_c \bar\psi_\phi(\bm{p}',a')\[\gamma^\mu \mathcal F_1(-q^2) + i\frac{\kappa}{2M}\mathcal F_2(-q^2) \sigma^{\mu\nu} q_\nu \] \psi_\phi(\bm{p},a)\nonumber\\
=&i e Z_c \bar\psi_\phi(\bm{p}',a')\[\frac{p^\mu+p'^\mu}{2M} \mathcal F_1(-q^2) + i\frac{\mathcal F_1(-q^2) +\kappa \mathcal F_2(-q^2)}{2M} \sigma^{\mu\nu} q_\nu \] \psi_\phi(\bm{p},a), 
\end{align}
that is rewritten 
using the Gordon identity in the last line.  The constant $\kappa$ is the anomalous magnetic moment and $Z_c$ the charge of the halo nucleus core.   For non-relativistic kinematics, in the Breit frame $(q_0=0,\bm{q})$, we get
\begin{align}\label{eq:EMforms}
i\Gamma^0 = & ie Z_c \bar u_\phi(\bm{p}',a')\mathcal F_1(|\bm{q}|^2) u_\phi(\bm{p},a), \\
i\Gamma^i =& ie  Z_c \bar u_\phi(\bm{p}',a')\[ \frac{p^i +p'^i}{2M} \mathcal F_1(|\bm{q}|^2) 
+i\frac{\mathcal F_1(|\bm{q}|^2) +\kappa\mathcal F_2(|\bm{q}|^2)}{2M}\epsilon^{ijk}\sigma_k q_j
\] u_\phi(\bm{p},a).\nonumber
\end{align}
In the Sach form that is commonly used for a physical interpretation, we write the charge $G_E(|\bm{q}|^2)$ and magnetic $G_M(|\bm{q}|^2)$ form factors as
\begin{align}
\label{eq:Sach}
G_E(|\bm{q}|^2) = &\mathcal F_1(|\bm{q}|^2)-\tau\kappa\mathcal F_2(|\bm{q}|^2),\\
G_M(|\bm{q}|^2)=&\mathcal F_1(|\bm{q}|^2) +\kappa \mathcal F_2(|\bm{q}|^2), \nonumber
\end{align} 
with $\tau=|\bm{q}|^2/(4 M^2)$. 
In the EFT the form factors $\mathcal F_i$s are $\mathcal O(1)$ in the $Q/\Lambda$ expansion as we show later in Section~\ref{sec_EFT}.  We count $|\bm{q}|\sim Q$ at low photon momentum exchange.  Thus the magnetic form factor $G_M$ gets contribution from both $\mathcal F_1$ and $\mathcal F_2$ whereas the electric form factor $G_E$ only gets contribution from $\mathcal F_1$ upto NLO.  The $\mathcal F_2$ term in $G_E$ is the so called Darwin-Foldy contribution.

The electric and magnetic form factors are normalized such that for small $|\bm{q}|$
\begin{align}
G_E(|\bm{q}|^2)\approx 1-\frac{1}{6} \langle r_E^2 \rangle|\bm{q}|^2+\cdots,
\end{align}
where $\sqrt{\langle r_E^2\rangle}$ is the charge radius and 
\begin{align}
\frac{eZ_c}{2M}G_M(|\bm{q}|^2)\approx\kappa_\phi\mu_N \[1-\frac{1}{6} \langle r_M^2\rangle |\bm{q}|^2+\cdots\],
\end{align}
where $\kappa_\phi$ is the halo nucleus  magnetic moment and $\sqrt{\langle r_M^2\rangle}$ the magnetic radius. 

The differential elastic scattering cross section in the laboratory frame  is written as
\begin{align}
\label{eq:diffcsection}
\frac{d\sigma}{d\Omega}={\frac{d\sigma}{d\Omega}} |_{Mott}  \[\mathcal A(|\bm{q}|^2)+ \mathcal B(|\bm{q}|^2) \tan^{2}\(\frac{\theta}{2}\)\] , 
\end{align}
where
\begin{align}
\label{eq:difcrosssection}
\mathcal A(|\bm{q}|^2)& ={\mathcal F_1^2(|\bm{q}|^2)}+\tau\kappa^2 \mathcal F_2^2(|\bm{q}|^2)
= \frac{1}{1+\tau}[G_E^2(|\bm{q}|^2) +\tau G_M^2(|\bm{q}|^2)], \\
\mathcal B(|\bm{q}|^2) &=2\tau[\mathcal F_1(|\bm{q}|^2)+\kappa \mathcal F_2(|\bm{q}|^2 )]^2 = 2\tau G_M^2(|\bm{q}|^2). \nonumber
\end{align}

\section{Effective Field Theories}\label{sec_EFT}

The halo nuclei $^{11}$Be, $^{15}$C and $^{19}$C ground states all have spin-parity assignment $\frac{1}{2}^+$. They are treated as a shallow bound state of a single neutron and a spin zero core in the $s$-wave.  This is reasonable as the binding energy of the ground state is much smaller than the energy needed to break the core or the excited state energies of the core~\cite{Phillips:2010dt,Hammer:2011ye,PhysRevC.86.044608,Acharya:2013nia}. The EFT calculations in these halo systems, so far,  agree with available data within the estimated theoretical errors.   The bound state is described by the strong interaction Lagrangian
\begin{align}\label{eq:Ls}
\mathcal L_s=\phi^\dagger_\alpha\[\Delta+iD_0+\frac{\bm{D}^2}{2M}\]\phi_\alpha+h\[\phi^\dagger_\alpha(N_\alpha C)+\operatorname{h.c.}\],
\end{align}
where $\phi_\alpha$ is an auxiliary field with spin index $\alpha$, $N_\alpha$ is the neutron field and $C$ is a scalar field for the core.  In the following we suppress the spin index. $D_\mu=\partial_\mu+ie Z_c A_\mu$ is the covariant derivative. The field $\phi$ represents the $\frac{1}{2}^+$ single neutron bound halo nucleus.  We take the neutron mass $M_n=939.6$ MeV, total mass $M=M_n+M_c$ where the core mass $M_c=9328$ MeV, 13044 MeV and 16792 MeV for $^{10}$Be, ${}^{14}$C and ${}^{18}$C respectively.   
The strong interaction couplings $\Delta$, $h$ is specific to the particular halo nucleus we consider and would in general be different from one system to the next. 

\begin{figure}[thb]
\begin{center}
\includegraphics[width=0.47\textwidth,clip=true]{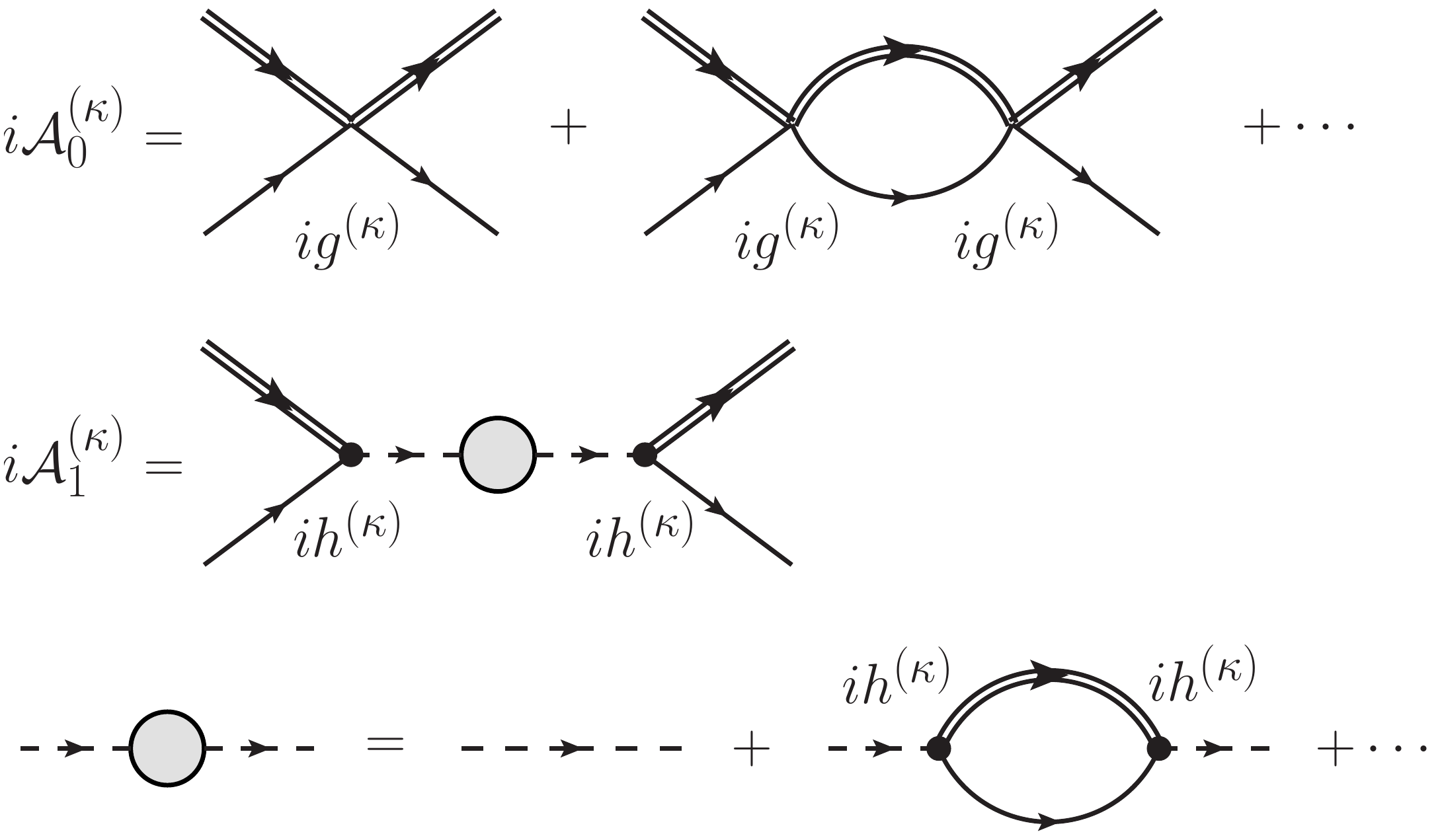} 
\end{center}
\caption{\protect Elastic scattering amplitudes $\mathcal A$ in $s$-wave. 
 Single line is the neutron propagator,  double line represent the dimer $\phi$  propagator, 
dashed line the bare dimer propagator. 
}
\label{fig:scattering}
\end{figure}

The EFT couplings are related to elastic scattering parameters. Calculating the elastic neutron-core scattering amplitude in Fig~\ref{fig:scattering}, we get 
\begin{align}\label{eq:AmpEFT}
i\mathcal A(p)=-i h^2 D_\phi(\frac{p^2}{2\mu},0)= -\frac{i h^2}{\Delta+p^2/(2\mu)+\mu h^2(\lambda+i p)/(2\pi)},
\end{align}
where the dressed $\phi$ propagator is
\begin{align}
i D_\phi(p_0,\bm{p})=&\frac{i}{\Delta+p_0-p^2/(2M)+i h^2 f_0(p_0,\bm{p})},\\
f_0(p_0,\bm{p})= &-i 2\mu\(\frac{\lambda}{2}\)^{4-D}\int 
\frac{d^{D-1}\bm{q}}{(2\pi)^{D-1}}\frac{1}{q^2- 2\mu p_0 +\mu p^2/M -i 0^+}\nonumber\\
=& 
-\frac{i\mu}{2\pi}(\lambda-\sqrt{-2\mu p_0 +\mu p^2/M-i 0^+}), \nonumber 
\end{align}
and $\lambda$ is the renormalization scale~\cite{Kaplan:1998tg} and $\mu=M_n M_c/(M_n+M_c)$ the reduced mass. Comparing the above relation to the equivalent one from the effective range expansion in the $s$-wave 
\begin{align}
i\mathcal A(p)=\frac{2\pi}{\mu}\frac{i}{p\cot\delta-i p}\approx \frac{2\pi}{\mu}\frac{i}{-\gamma+\rho(p^2+\gamma^2)/2-i p},
\end{align}
we get
\begin{align}\label{eq:EFTcouplings}
\frac{2\pi\Delta}{\mu h^2}+\lambda=&\gamma-\frac{1}{2}\rho\gamma^2,\\
-\frac{2\pi}{h^2\mu^2}=&\rho. \nonumber
\end{align}
The binding momentum $\gamma$ is determined from the binding energy $B=\gamma^2/(2\mu)$.  The effective range $\rho$ is typically less constrained from data as elastic neutron scattering data is scarce. However, $\rho$ can be constrained from radiative capture or Coulomb dissociation data when available. In the EFT power counting $\gamma\sim Q$ for shallow bound states and contributes at leading order.  \emph{A priori} it is not known how $\rho$ that has dimensions of length should scale. If $\rho\sim 1/\Lambda$ it is a next-to-leading order effect whereas if $\rho\sim 1/Q$ it contributes at leading order.  We consider both the situations later -- perturbative $\rho$ for $^{11}$Be and $^{19}$C and non-perturbative $\rho$ for $^{15}$C.

The form factor calculations also depend on the magnetic moment coupling of the neutron and possible two-body currents.  We consider the following magnetic operators in addition to the interactions in Eq.~(\ref{eq:Ls}): 
\begin{align}
O_{EM}=2\kappa_n\mu_NN^\dagger\bm\sigma\cdot\bm B N +\mu_N L_M\phi^\dagger\bm\sigma\cdot\bm B\phi,
\end{align}
where $\kappa_N=-1.91304$ is the neutron anomalous magnetic  moment, $\mu_N$ the nuclear magneton. $L_M$ is the dimensionless  couplings for a magnetic two-body current.

\begin{figure}[thb]
\begin{center}
\includegraphics[width=0.7\textwidth,clip=true]{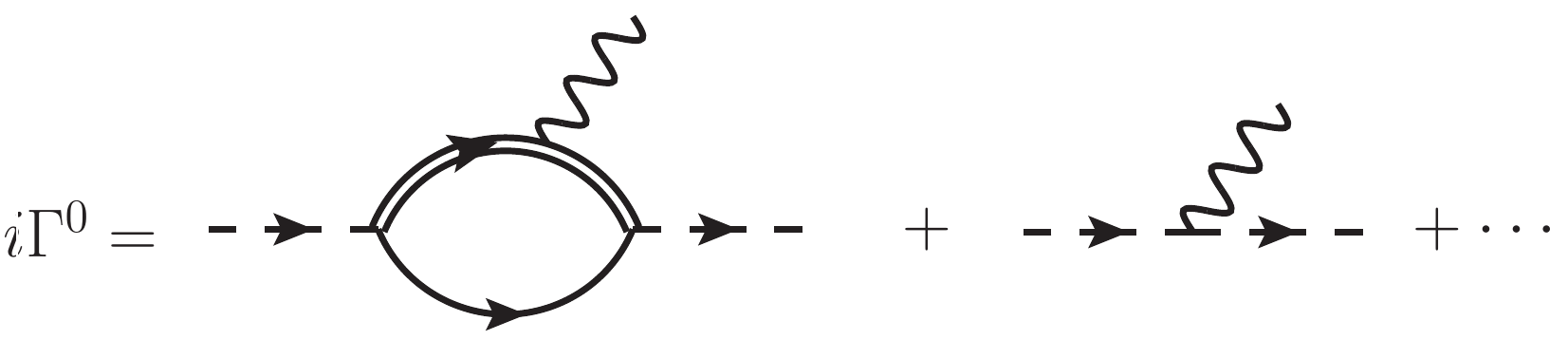} 
\end{center}
\caption{\protect EFT calculation of $\Gamma^0$. The wavy lines correspond to $A_0$ photons.}
\label{fig:GammaZero}
\end{figure}

The form factors in Eq.~(\ref{eq:EMforms}) are calculated from Feynman diagrams with a photon between initial and final ground state $\phi$ with momenta $\bm{p}$ and $\bm{p}'$, respectively.  In general this requires initial and final state interaction description of the ground state, and  electromagnetic current insertion in intermediate state. 
The EFT calculation of $\Gamma^0$ corresponds to the diagrams in Fig.~\ref{fig:GammaZero} where a $A_0$ photon is inserted between the initial and final ground state.  We get
\begin{align}\label{eq:Gamma0}
i\Gamma^0 = &i e Z_c Z_\phi \bar u_\phi(\bm{p'},a')\[  h^2\frac{\mu M_c}{\pi |\bm{q}|} \tan^{-1}\(\frac{\mu|\bm{q}|}{2 M_c\gamma}\) +1 \]u_\phi(\bm{p},a), 
\end{align}
where the first term is the contribution from the one-loop diagram and the second term is from the tree-level diagram.  The overall factor $Z_\phi$ is the wavefunction renormalization  that is defined as the residue of the dimer $\phi$ propagator at the bound state energy pole~\cite{Chen:1999tn}
\begin{align}
Z_\phi^{-1} =\frac{\partial}{\partial p_0}[D_\phi(p_0,\bm{p})]^{-1}\Big|_{p_0=p^2/(2M)-B} = 1+\frac{\mu^2 h^2}{2\pi\gamma}
=-\frac{1-\rho\gamma}{\rho\gamma}.
\end{align}
We used the relation $B=\gamma^2/(2\mu)$ for the shallow bound nucleus and $h^2=-2\pi/(\rho\mu^2)$ from before.

For the halo nuclei  $^{11}$Be and $^{19}$C  in 
Section~\ref{sec_results},  $\rho\sim 1/\Lambda$ and we see that the second term in Eq.~(\ref{eq:Gamma0}) is  $\mathcal O(Q/\Lambda)$ smaller compared to the first term. Though the effective range correction contributes at NLO, some of the $\rho$'s are large Refs.~\cite{Hammer:2011ye,Fernando:2011ts,Acharya:2013nia} which motivates us to use the ``zed"-parameterization~\cite{Phillips:1999hh}. In this parameterization, the wave function renormalization is reproduced exactly at NLO.   For the halo nuclei $^{15}$C, $\rho\sim 1/Q$ and we keep the effective range contributions exactly by treating it non-perturbatively. In this case, both the terms in Eq.~(\ref{eq:Gamma0}) contribute at the same order. We start with a description of the zed-parameterization.

A convenient starting point for formulating the zed-parameterization with dimers is Eq.~(\ref{eq:EFTcouplings}). 
The EFT power counting assumption  $\gamma\sim\lambda\sim Q$ and $\rho\sim 1/\Lambda$ implies $\Delta\sim Q$, $h^2\sim1/\Lambda$.  To renormalize the theory systematically,  we expand the couplings $\Delta$ and $h$ as
\begin{align}
\Delta &=\Delta_1+\Delta_2+\Delta_3+\cdots,\\
h &=h_0+h_1+h_2+\cdots, \nonumber
\end{align}
where the subscript indicates the scaling with the powers of $Q$ in the $Q/\Lambda$ expansion. Then by inspection of Eq.~(\ref{eq:AmpEFT}), one sees that $\Delta_1$ and $h_0$ along with the unitary cut contribution $ip$ contributes at LO while the $p^2/(2\mu)$ piece associated with the effective range expansion would appear at higher order.  As the combination $h^2 Z_\phi= 2\pi\gamma/[\mu^2(1-\rho\gamma)]$ enters the calculation often, in the zed-parameterization we rewrite $1/(1-\rho\gamma)$ as $1+(Z_d-1)$ where $Z_d-1$ is treated as order $Q/\Lambda$.  
Consistently applying the $\Delta_n$, $h_n$ expansion to Eq.~(\ref{eq:EFTcouplings}), then yields 
\begin{align}
&\Delta_1=-\frac{\gamma(\gamma-\lambda)}{\mu(Z_d-1)},& &\Delta_2=-\frac{\gamma(\gamma-2\lambda)}{2\mu},\\
&h_0^2=-\frac{2\pi\gamma}{\mu^2(Z_d-1)},&  &h_1^2=-\frac{\pi\gamma(Z_d-1)}{2\mu^2}, \nonumber\\
&\hspace{0.25in} \vdots & &\hspace{0.25in}\vdots{} \nonumber
\end{align} 
where the perturbative expansion for $\Delta$ beyond the terms shown vanish but for $h^2$ continues. It is straightforward then to show that 
\begin{align}
i\mathcal A(p)=\frac{2\pi}{\mu}\frac{i}{-\gamma-ip}[1+(Z_d-1)+0+0+0+\cdots]\ ,
\end{align}
as derived in Ref.~\cite{Phillips:1999hh}. $\rho$ and $Z_d$ are related in perturbation as $Z_d=1+\rho\gamma+(\rho\gamma)^2+\cdots$. We express physical observables in terms of $Z_d-1$ instead of $\rho$ when the perturbative expansion is valid.  In situations where $\rho\sim1/Q$, $\rho\gamma\sim 1$, we do not treat $Z_d-1$ as a perturbation, and $Z_d=1/(1-\rho\gamma)$ is not expanded.

\begin{figure}[thb]
\begin{center}
\includegraphics[width=0.7\textwidth,clip=true]{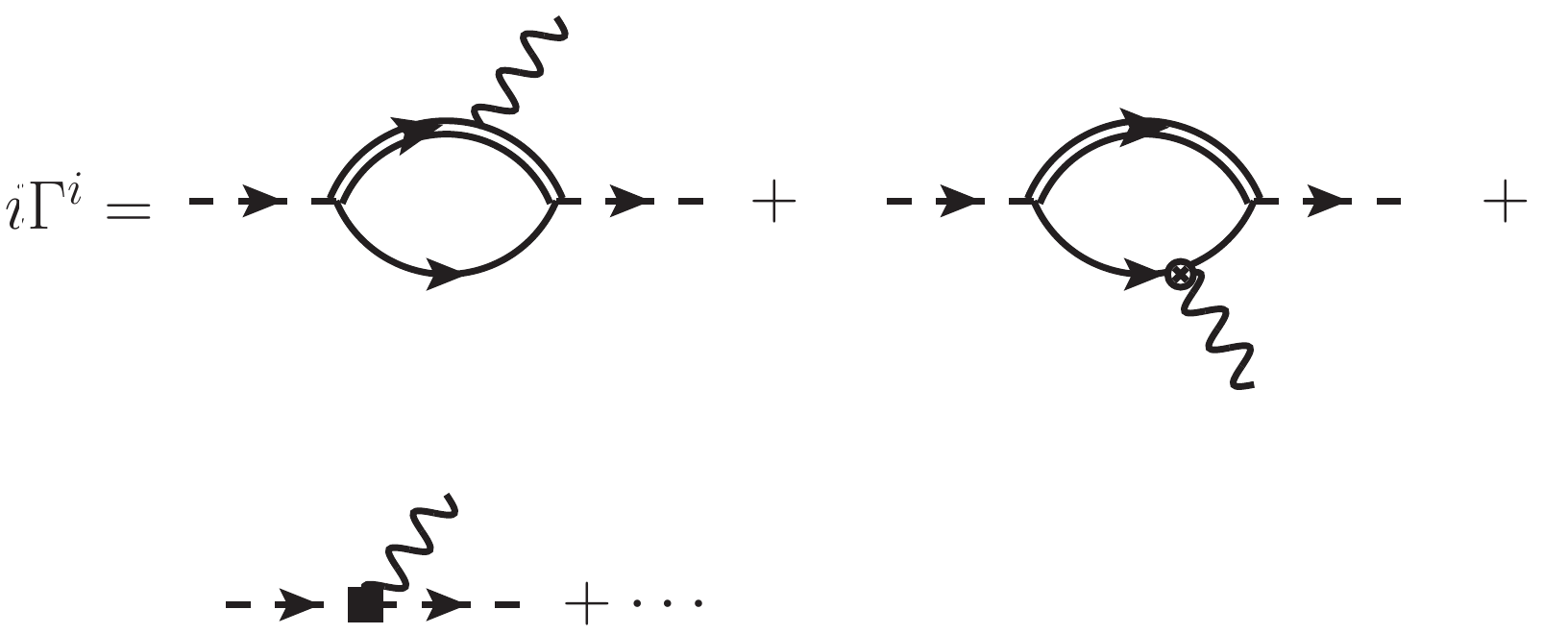} 
\end{center}
\caption{\protect EFT calculation of $\Gamma^i$. The wavy lines correspond to $A_i$ photons. The magnetic coupling is represented in the second diagram with a  $\otimes$, and the two-body current in the third diagram is represented with a filled square. }
\label{fig:GammaI}
\end{figure}
The contribution to $\Gamma^i$ follows from the diagrams in Fig.~\ref{fig:GammaI} that gives:
\begin{align}\label{eq:GammaI}
i\Gamma^i=&ie Z_c Z_\phi \bar u_\phi(\bm{p},a)\left\{
\frac{p_i +p'_i}{2M} \[ h^2\frac{\mu M_c}{\pi |\bm{q}|} \tan^{-1}\(\frac{\mu|\bm{q}|}{2 M_c\gamma}\) +1
\] 
\right. \\ &\left.
+i\frac{\mu_N}{e Z_c}\[ {h^2 \kappa_n}\frac{\mu M_n}{\pi |\bm{q}|} \tan^{-1}\(\frac{\mu|\bm{q}|}{2 M_n\gamma}\) +L_M
\]\epsilon^{ijk}\sigma_j q_k
\right\}u_\phi(\bm{p}',a').\nonumber
\end{align}
$\Gamma^i$ receives contribution from magnetic photons $A_i$  that includes contribution from both magnetic moment coupling and the electromagnetic current generated by the orbital motion of the charged $^{10}$Be, $^{14}$C or $^{18}$C core. The two-body current contribution indicated by the dimensionless coupling 
$L_M$ which is assumed to have a natural size $\mathcal O(1)$.  The two-body current contributes at NLO for perturbative $\rho$ and at LO for non-perturbative 
$\rho$ in the $Q/\Lambda$ expansion. We first derive some expression for pertrubative $\rho$ where we apply the zed-parameterization. We consider the non-perturbative case later separately when we discuss the $^{15}$C nucleus in Section~\ref{sec_results} to keep the discussion simple.

Comparing Eqs. ~(\ref{eq:EMforms}), (\ref{eq:Gamma0}) and  (\ref{eq:GammaI}), we get
\begin{align}
\mathcal F_1(|\bm{q}|^2)&=Z_\phi \[  h^2\frac{\mu M_c}{\pi |\bm{q}|} \tan^{-1}\(\frac{\mu|\bm{q}|}{2 M_c\gamma}\) +1 \] \\
&=\frac{2 M_c\gamma}{\mu |\bm{q}|} \tan^{-1}\(\frac{\mu|\bm{q}|}{2 M_c\gamma}\)
+(Z_d-1)\[  \frac{2 M_c\gamma}{\mu |\bm{q}|} \tan^{-1}\(\frac{\mu|\bm{q}|}{2 M_c\gamma}\) -1\], \nonumber
\\
G_M(|\bm{q}|^2)&=\frac{2 M\mu_N}{e Z_c}  Z_\phi\[ \frac{h^2 g_n}{2}\frac{\mu M_n}{\pi |\bm{q}|} \tan^{-1}\(\frac{\mu|\bm{q}|}{2 M_n\gamma}\) +L_M
\] \nonumber \\
&=\frac{2 M\mu_N}{e Z_c}\frac{2 M_n\gamma\kappa_n}{\mu |\bm{q}|} \tan^{-1}\(\frac{\mu|\bm{q}|}{2 M_n\gamma}\)\nonumber\\
&+(Z_d-1)\frac{2 M\mu_N}{e Z_c}\[  \frac{2 M_n\gamma\kappa_n}{\mu |\bm{q}|} \tan^{-1}\(\frac{\mu|\bm{q}|}{2 M_n\gamma}\) -L_M\], \nonumber
\\
G_E(|\bm{q}|^2)&=(1+\tau )\mathcal F_1(|\bm{q}|^2)-\tau G_M(|\bm{q}|^2). \nonumber
\end{align}
The form factor $\mathcal F_1$ can be determined once the binding momentum $\gamma$ and wave function renormalization constant $Z_d-1$ is known. For the magnetic form factor $G_M$, the two-body coupling $L_M$ is also need.

Expanding the electric form factor in $|\bm q|$ followed by an expansion in $Z_d-1\sim Q/\Lambda$, we get
$\rho\gamma\sim Q/\Lambda$ , we get to NLO
\begin{align}
G_E(|\bm{q}|^2) &\approx 1-\frac{\mu^2}{12M_c^2\gamma^2}[1+(Z_d-1)] |\bm q|^2.
\end{align}
The charge normalization at low $|\bm q|^2$ is as expected. 
In the electric form factor we ignored the Darwin-Foldy contributions which appear at higher order. In the EFT the core of the halo nucleus is treated as point-like.  However, to compare the charge radius with experimental values one has to add the finite charge radius of the core in quadrature. We write
\begin{align}
&\langle r_E^2\rangle\approx \frac{\mu^2}{2 M_c^2 \gamma^2}[1+(Z_d-1)]+\langle r_c^2\rangle, 
\end{align}
expanded to NLO  where $\sqrt{\langle r_c^2\rangle}$ is the core charge radius. The LO charge radius is entirely determined by the halo nucleus binding energy. The effective range parameter contributes at NLO which we discuss in the next section.

For the magnetic form factor we get
\begin{align}
\frac{e Z_c}{2M}G_M(|\bm{q}|^2)&\approx \mu_N[\kappa_n+(Z_d-1)(\kappa_n-L_M)]
-\frac{\mu^2}{12 M_n^2\gamma^2}\kappa_n\mu_N[1+(Z_d-1)]|\bm q|^2,
\end{align}
expanded to NLO for small $|\bm q|^2$.  The halo nuclei magnetic moment is identified as
\begin{align}\label{eq:kappaEFT}
\kappa_\phi=\kappa_n+(Z_d-1)(\kappa_n-L_M),
\end{align}
where the LO result is just the Schmidt value associated with the magnetic moment of the valance neutron. The LO magnetic radius is in analogy to the charge radius given by
\begin{align}
\langle r_M^2\rangle\approx \frac{\mu^2}{2 M_n^2 \gamma^2}[1+(Z_d-1)] .
\end{align}

From the above analysis that is applicable to perturbative $\rho$, we see that the LO result is known from the binding energy and the neutron magnetic moment. At NLO, contribution from both the effective range and a two-body current is needed to determine the electromagnetic form factors. 

\section{Form Factors}\label{sec_results}

In this section  we apply the expressions derived above to ${}^{11}$Be, ${}^{15}$C and ${}^{19}$C nuclei, and calculate the corresponding electromagnetic form factors.

\subsection{$^{11}$Be}
The $s$-wave $\frac{1}{2}^+$ state and the $p$-wave state $\frac{1}{2}^-$ of ${}^{11}$Be was analyzed in Ref.~\cite{Hammer201117}.  We only consider the $s$-wave $\frac{1}{2}^+$ state  here.  In the EFT power counting with $\gamma\sim |\bm q|\sim Z_d-1\sim Q$ ,
 the LO and NLO contributions to the form factors $\mathcal A$ is:
 \begin{align}\label{eq:AZd}
 \mathcal A(|\bm q|^2)\approx \frac{4 M^2_c\gamma^2}{\mu^2 |\bm q|^2}\[\tan^{-1}\(\frac{\mu|\bm q|}{2M_c \gamma}\)\]^2
 +(Z_d-1) &\frac{4 M_c\gamma}{\mu |\bm q|}\tan^{-1}\(\frac{\mu|\bm q|}{2M_c \gamma}\)   \\
 &\times\[\frac{2 M_c\gamma}{\mu |\bm q|}\tan^{-1}\(\frac{\mu|\bm q|}{2M_c \gamma}\)-1\],  \nonumber
 \end{align}
 Up to NLO, $\mathcal A$ depends only on the electric form factor $G_E$. 
The binding momentum in the above relation is determined from the valance neutron separation energy $B=500$ keV as $\gamma=\sqrt{2\mu B}\approx 29.22$ MeV~\cite{Kelley1990}.  In Ref.~\cite{Hammer201117} the wave function normalization factor is determined from the 
 Coulomb dissociation of the $\frac{1}{2}^+$ state to neutron and 
${}^{10}$Be through E1 transition that determines $Z_d-1=0.69$. This corresponds to $\rho\gamma\approx 0.4$ in $Z_d=1/(1-\rho\gamma)$. Effective range corrections though perturbative are significant justifying the use of the zed-parameterization.  
 This yields a EFT charge radius $\langle r_E^2\rangle^{1/2}=(2.40\pm0.02)$ fm using the experimental ${}^{10}$Be radius  $\langle r_c^2\rangle^{1/2}=(2.357\pm0.018)$ fm that compares well with the experimental value $\langle r_E^2\rangle_\mathrm{exp}^{1/2}=(2.463\pm0.016)$ fm. 
The EFT expansion for this system is estimated to be around $Q/\Lambda\sim 0.3$-0.4. The NLO result  is expected to have an error of about 10-15\% from the NNLO $(Q/\Lambda)^2$ corrections.  The final state interaction in the $p$-wave that is treated perturbatively for natural sized parameters contribute at NNLO.

\begin{figure}[thb]
\begin{center}
\includegraphics[width=0.7\textwidth,clip=true]{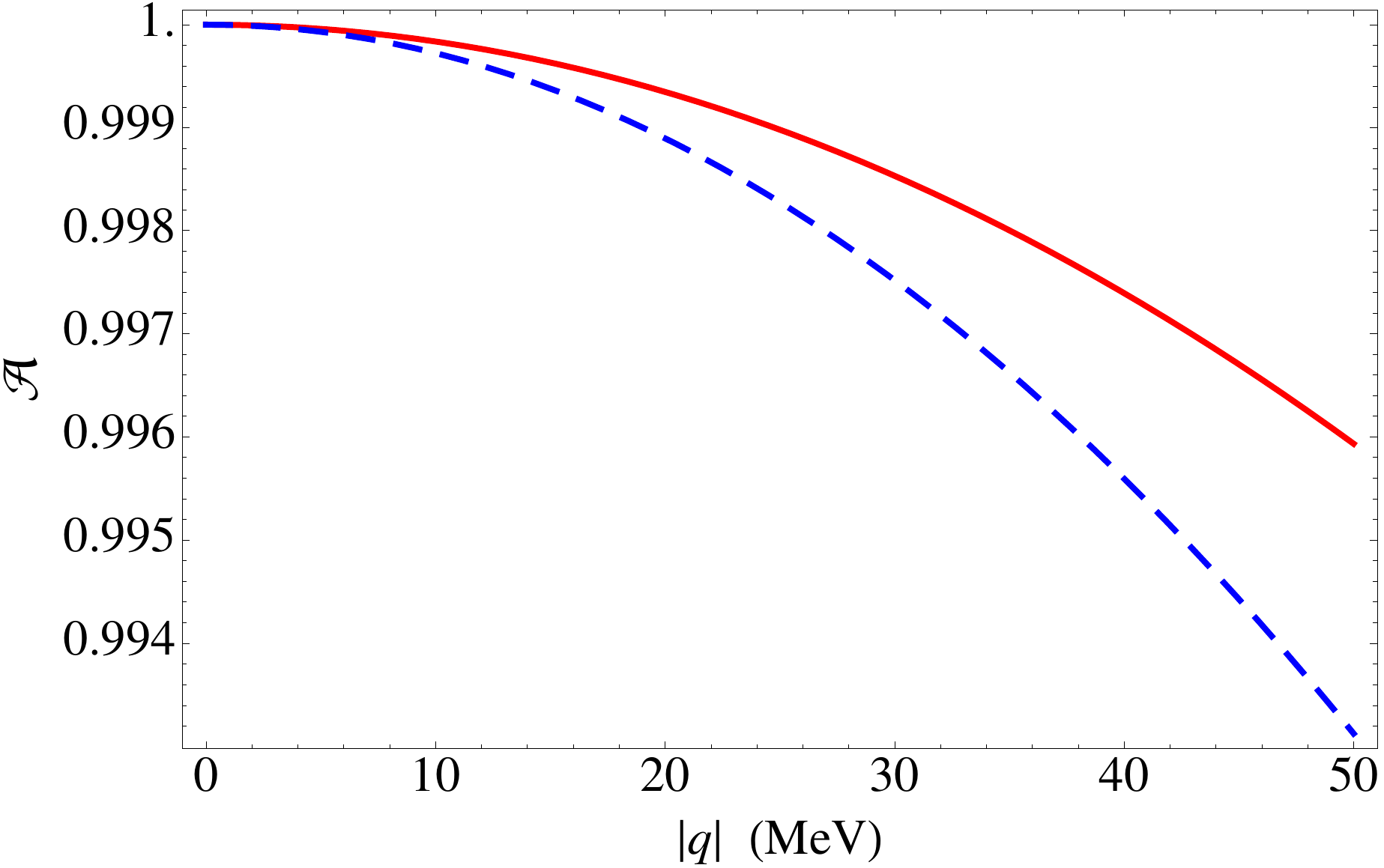}  \includegraphics[width=0.7\textwidth,clip=true]{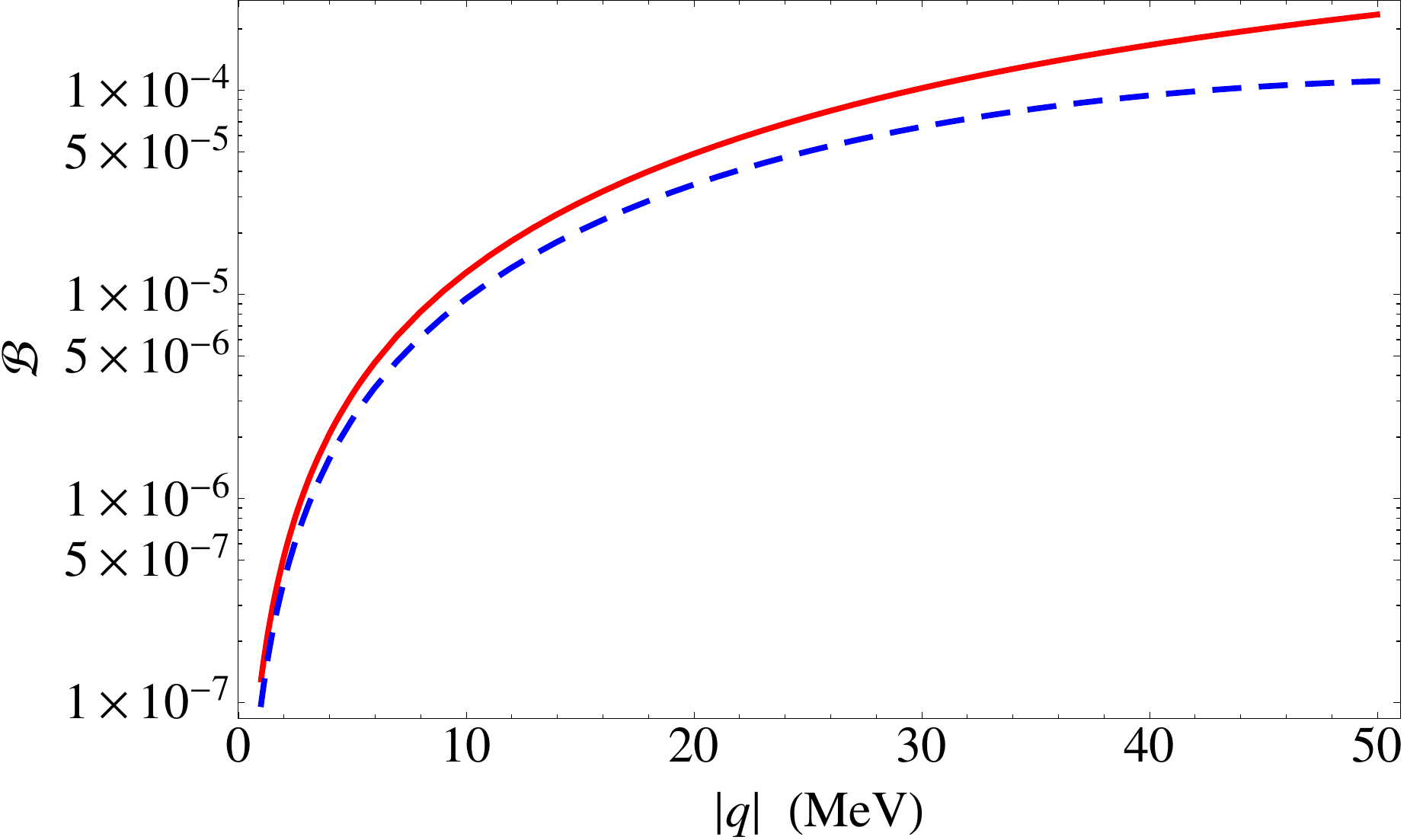} 
\end{center}
\caption{\protect  Form factors for $^{11}$Be. Solid red curve LO contribution; and dashed blue curve LO + NLO contributions. }
\label{fig:ABbe11}
\end{figure}

The form factor $\mathcal B$ can be expanded similarly to get 
\begin{align}\label{eq:BZd}
 \frac{e^2 Z_c^2}{4M^2\mu_N^2}
 \mathcal B(|\bm q|^2)&\approx
 \kappa_n^2\frac{2 M_n^2\gamma^2}{M^2\mu^2 }\[\tan^{-1}\(\frac{\mu|\bm q|}{2M_n \gamma}\)\]^2 \\
 &+ (Z_d-1)\kappa_n\frac{2 M_n\gamma|\bm q|}{M^2 \mu}\tan^{-1}\(\frac{\mu|\bm q|}{2M_n \gamma}\)
\[\frac{2 M_n\gamma \kappa_n}{\mu |\bm q|}\tan^{-1}\(\frac{\mu|\bm q|}{2M_n \gamma}\) -L_M\]. \nonumber
 \end{align}
 To determine $\mathcal B$ at NLO we need to know $L_M$ besides $\gamma$ and $Z_d$. We fit $L_M$ 
 to the known magnetic moment for $^{11}$Be, 
$\kappa_\phi^{(exp)}=-1.6814$~\cite{PhysRevLett.102B}, which gives $L_M=-2.25$ from Eq.~(\ref{eq:kappaEFT}).  This is a reasonable value for a dimensionless coupling in the EFT power counting where we assumed it to be $\mathcal O(1)$. 
In Fig.~\ref{fig:ABbe11} we plot the form factors $\mathcal A(|\bm q|^2)$ and $\mathcal B(|\bm q|^2)$.  

\subsection{$^{19}$C}
The halo nuclei $^{19}$C was considered in halo EFT in Ref.~\cite{Acharya:2013nia}. The authors calculated the radiative capture $^{18}$C$(n,\gamma)^{19}$C 
and breakup $^{19}$C$(\gamma,n)^{18}$C cross  section. The EFT analysis extracts the binding energy as $0.575\pm 0.055$ MeV and $Z_d-1\approx 0.73$, where we only indicate the central value for $Z_d$ that enters at NLO.

\begin{figure}[thb]
\begin{center}
\includegraphics[width=0.7\textwidth,clip=true]{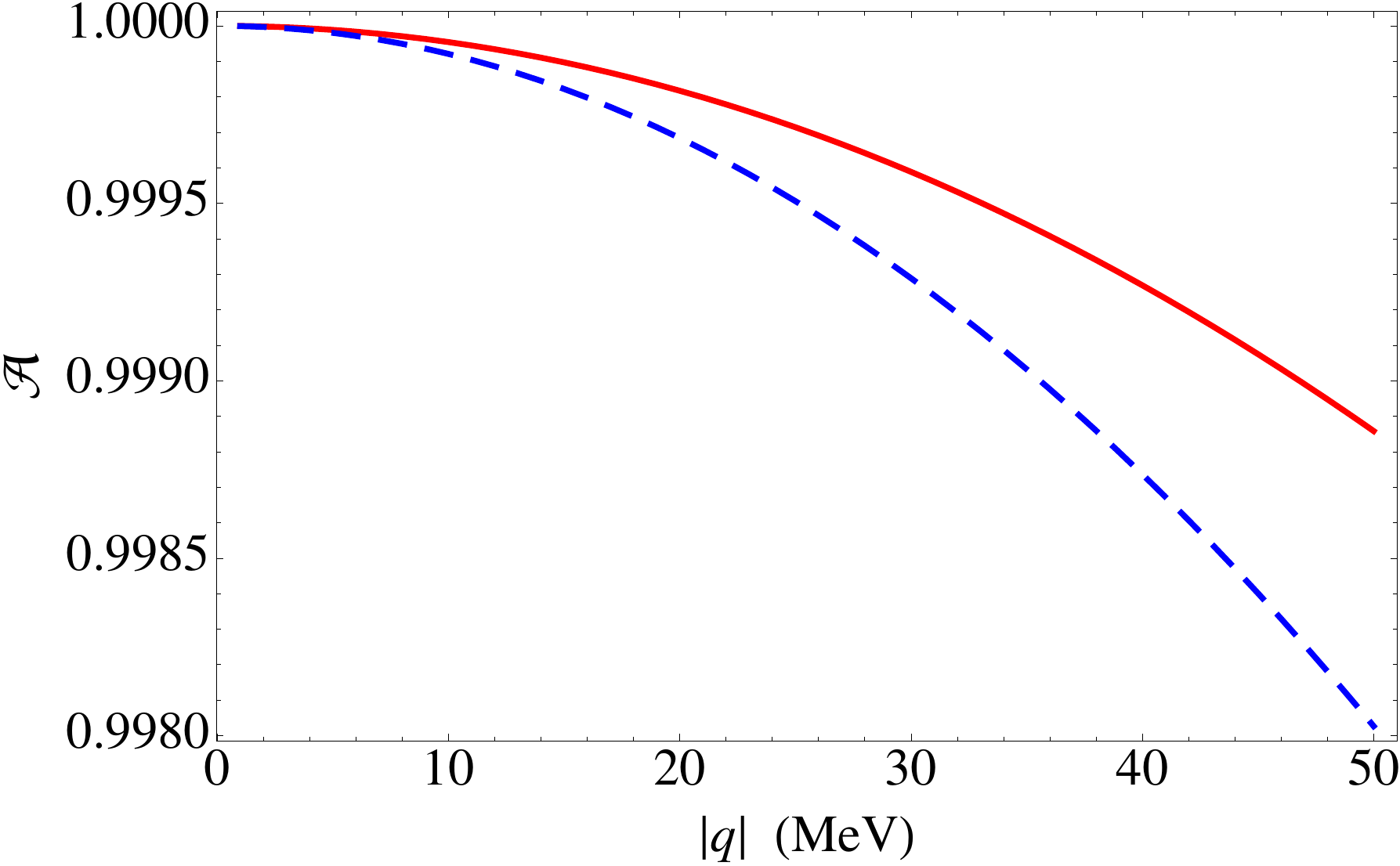}  \includegraphics[width=0.7\textwidth,clip=true]{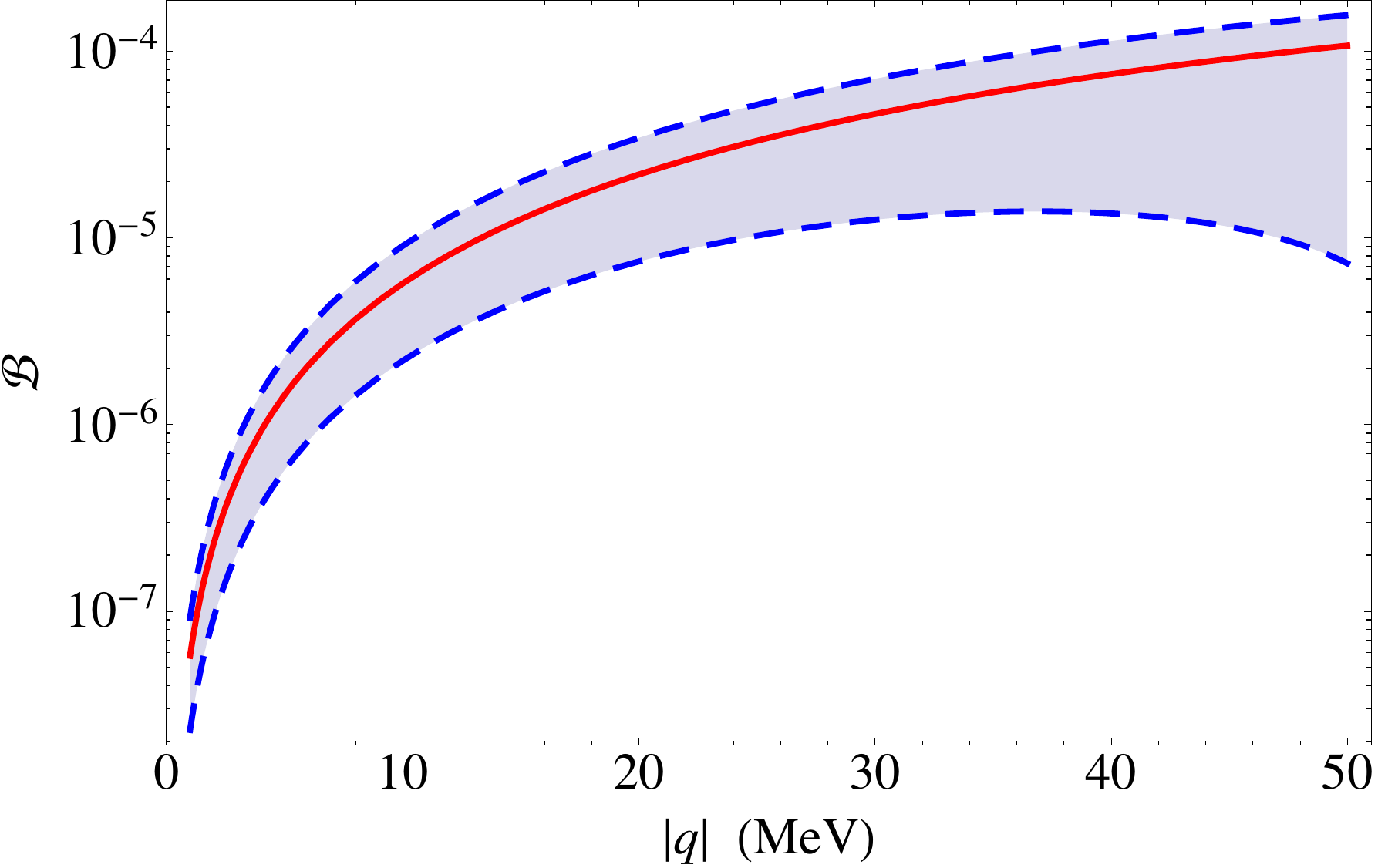} 
\end{center}
\caption{\protect  Form factors for $^{19}$C. Solid red curve LO contribution; and dashed blue curves LO + NLO contributions.  The shaded area between the blue curves indicate a range of NLO values  as explained in the text.}
\label{fig:ABc19}
\end{figure}

The charge radius and the magnetic moment for $^{19}$C are not known.  The analysis for this system is similar to the $^{11}$Be system. We can make a NLO prediction for the charge radius in halo EFT
\begin{align}
\langle r_E^2\rangle-\langle r_c^2\rangle
\approx \frac{\mu^2}{2 M_c^2 \gamma^2}[1+(Z_d-1)]=0.0534\times[1+0.73]\ \mathrm{fm}^2\approx 0.09\ \mathrm{fm}^2,
\end{align}
where we used the central values for the parameters. The form factor $\mathcal A$ is plotted in Fig.~\ref{fig:ABc19} using Eq.~(\ref{eq:AZd}). 

The magnetic moment at NLO depends on the two-body coupling $L_M$ that is not known. If we require that the NLO result for the magnetic moment be 
within 30\% of the LO Schmidt value, then 
we can estimate $-2.7\lesssim L_M\lesssim -1.1$. With this assumption we plot the form factor $\mathcal B$ in Fig.~\ref{fig:ABc19} using Eq.~(\ref{eq:BZd}). The shaded band indicates where the NLO result in halo EFT is expected to lie. The lower dashed blue curve corresponds to $L_M=-2.7$ and the upper dashed blue curve corresponds to $L_M=-1.1$. 
 
\subsection{$^{15}$C}

In Ref.~\cite{PhysRevC.86.044608}, $^{15}$C was treated as a single neutron halo nucleus with a $^{14}$C core. The radiative capture $^{14}$C$(n,\gamma)^{15}$C and breakup $^{15}$C$(\gamma,n)^{14}$C processes (through Coulomb dissociation) were calculated. The capture process proceeds through E1 transition from an initial $^2P_{1/2}$ and $^2P_{3/2}$ state to $^2S_{1/2}$ final state.  The breakup process is related to the capture through detailed balance.  The available direct capture and Coulomb dissociation data suggested that either the effective range or the $p$-wave interaction is non-perturbative. Here we revisit that discussion and present another analysis.  

\begin{figure}[thb]
\begin{center}
\includegraphics[width=0.7\textwidth,clip=true]{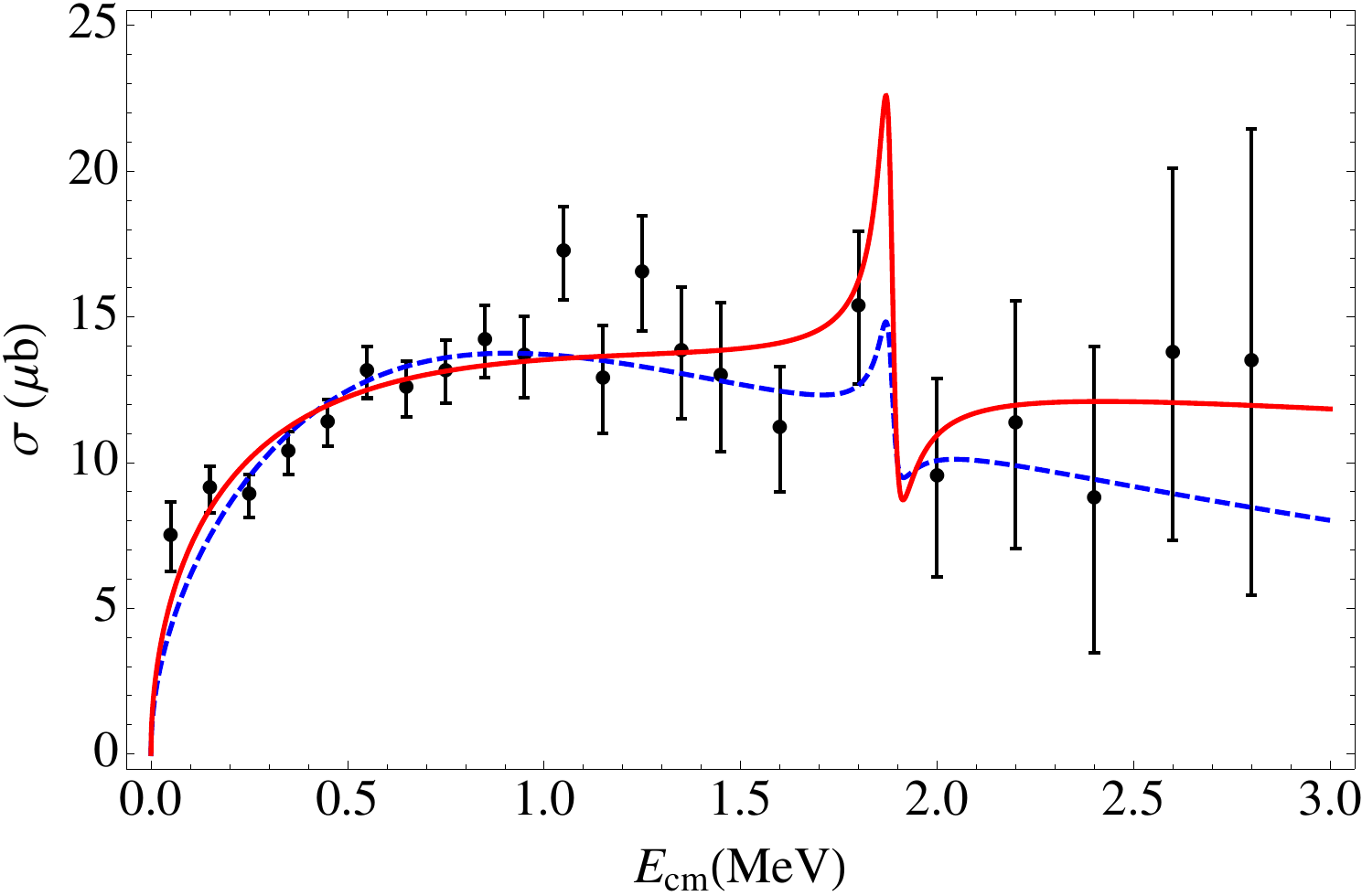}  
\end{center}
\caption{\protect  Capture cross section $^{14}$C$(n,\gamma)^{15}$C. The data is from Ref.~\cite{Nakamura2009}.  The  dashed  blue curve is a fit with non-perturbative $^2P_{3/2}$ interaction and the solid red curve is a fit with non-perturbative $^2S_{1/2}$ effective range interaction. The curves are fitted to c.m. energy 1.5 MeV as explained in the text. }
\label{fig:c15Capture}
\end{figure}

In the $^2P_{1/2}$  channel there is a resonance at  energy $E_r\approx 1.885$ MeV with a width of about $\Gamma_r\approx 40$ keV. The $p$-wave scattering volume and effective range are fixed as~\cite{Fernando:2011ts}
\begin{align}
a_1=-\frac{\mu\Gamma_r}{p_r^5}\approx -5.6\times10^{-8}\ \mathrm{MeV}^{-3},\ \ \mathrm{and}\ \ r_1=-\frac{2p_r^3}{\mu\Gamma_r}\approx-11\times 10^{3}\ \mathrm{MeV}.
\end{align}   
Though these values are not fine tuned, near the resonance their contribution is kinematically enhanced~\cite{Bertulani:2002sz,Bedaque:2003wa}.  Away from the resonance, $p$-wave interaction in this channel is suppressed as expected.  The capture and Coulomb dissociation data then suggest that either the $^2S_{1/2}$ 
effective range $\rho$ (or $Z_d-1$) or the $^2P_{3/2}$ interaction is fine tuned to be non-perturbative. 
Unlike in Ref.~\cite{PhysRevC.86.044608} where the $^2P_{3/2}$ interaction was take as non-perturbative, a non-perturbative $s$-wave effective range $\rho$ gives a slightly better fit reducing the $\chi^2$ per degree of freedom from 1.70 to 1.26.  Qualitatively the more important difference is that whereas the non-perturbative $p$-wave interaction required two operators to be fine tuned in Ref.~\cite{PhysRevC.86.044608}, a fine tuned $\rho$ involves a single $s$-wave fine tuned operator. In Fig.~\ref{fig:c15Capture} we show the two fits to data for the capture process.  The dependence on the effective rage $\rho$ enters as a factor of $1/(1-\rho\gamma)$ where we do not expand in $\rho$. 
We find $\rho=2.67$ fm or $Z_d=2.66$ from the fit. For $^{15}$C, with a binding energy $B=1.2181$ MeV, $\rho\gamma\sim 0.6$ which makes effective range corrections  large. For this halo system, we treat $\rho\sim 1/Q$ and at LO we get for the charge radius
 \begin{align}
 \langle r_E^2\rangle-\langle r_c^2\rangle=\frac{\mu^2}{2M_c^2\gamma^2}\frac{1}{1-\rho\gamma}
 \approx 0.11\ \mathrm{fm}^2, 
 \end{align}
and for the magnetic moment
\begin{align}
\kappa_\phi=\frac{\kappa_n-\rho\gamma L_M}{1-\rho\gamma}. 
\end{align}
Experimentally only the magnitude of the magnetic moment is known as $\kappa_\phi^{(exp)}=(1.720\pm0.009)$~\cite{Asahi200288}. Assuming a shell-model configuration with a valence $s$-wave neutron dominating  the $^{15}$C ground state wave function with 97-98\% probability, a tentative experimental value  $\kappa_\phi^{(exp)}=-(1.77\pm0.05)$ ~\cite{Asahi200288} was extracted. From this we can extract $L_M=-2.0$.

\begin{figure}[thb]
\begin{center}
\includegraphics[width=0.7\textwidth,clip=true]{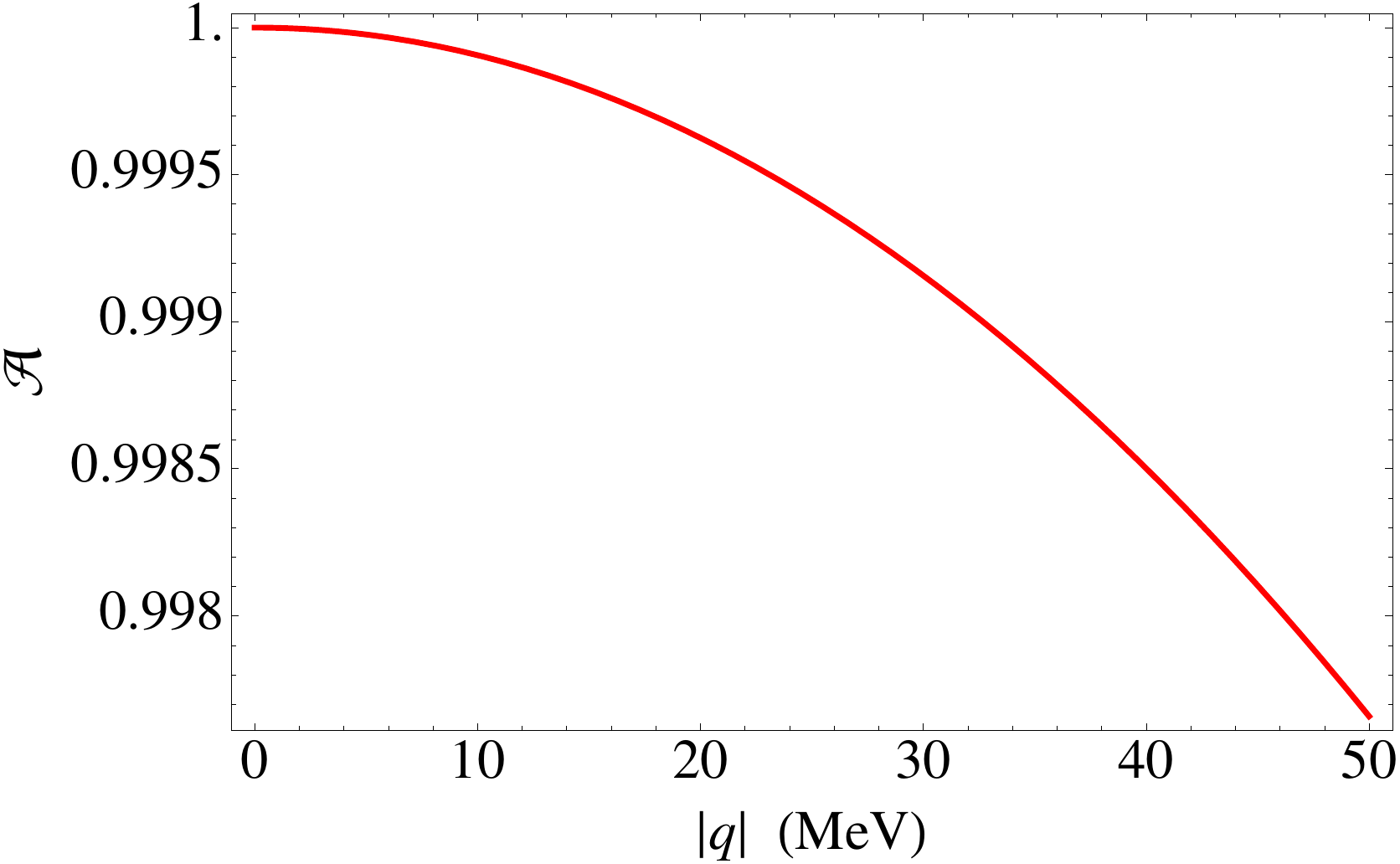}  \includegraphics[width=0.7\textwidth,clip=true]{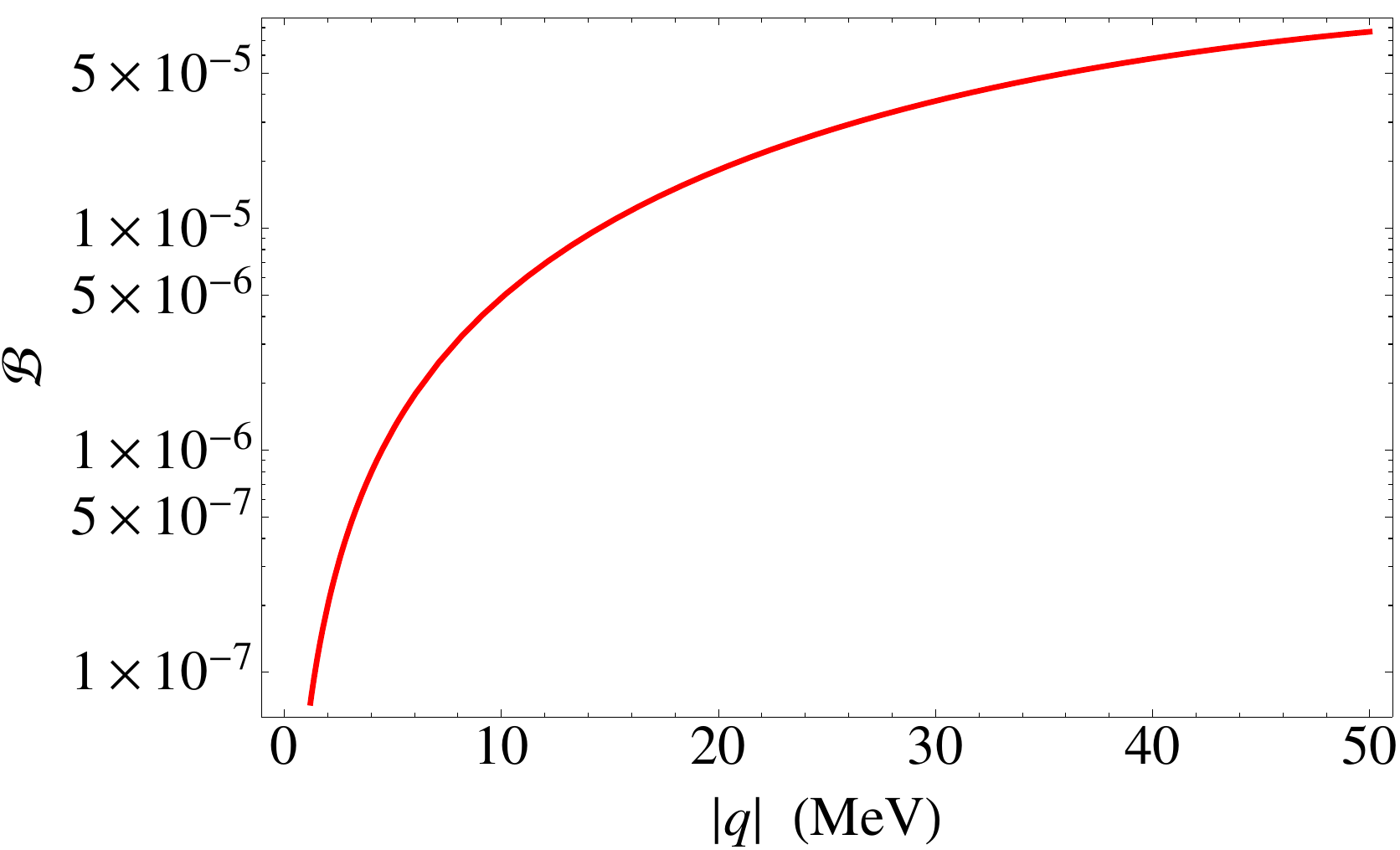} 
\end{center}
\caption{\protect  Form factors for $^{15}$C. Soldi red  curve LO contribution.}
\label{fig:ABc15}
\end{figure}

The form factors $A(|\bm q|^2)$ and $B(|\bm q|^2)$ for $^{15}$C is very similar to $^{11}$Be above except we do not expand in the effective range $\rho$ or equivalently in $Z_d-1$.  We get: 
\begin{align}
\mathcal A(|\bm q|^2)&=\frac{1}{(1-\rho\gamma)^2}\left[
\frac{2 M_c\gamma}{\mu |\bm q|}\tan^{-1}\(\frac{\mu|\bm q|}{2 M_c\gamma}\)-\rho\gamma
\right]^2,\\
\frac{e^2 Z_c^2}{4 M^2\mu_N^2}\mathcal B(|\bm q|^2)&= \frac{|\bm q|^2}{2 M^2} \frac{1}{(1-\rho\gamma)^2}
\left[\kappa_n\frac{2 M_n\gamma}{\mu |\bm q|}\tan^{-1}\(\frac{\mu |\bm q|}{2 M_n\gamma}\)-\rho\gamma L_M
\right]^2 .
\nonumber
\end{align}
The LO results are shown in Fig.~\ref{fig:ABc15}. 

\section{Conclusions}\label{sec_conclusions}
The electromagnetic form factors for several spin $\frac{1}{2}$ halo nuclei -- $^{11}$Be, $^{15}$C and $^{19}$C -- were calculated. The form factors probe the charge and magnetic distribution of the halo systems.  The calculations were performed using halo EFT where the halo nuclei is approximated as a single neutron bound to a nuclear core. We calculated the form factors to NLO except for $^{15}$C where the calculation was LO.  The form factors depended on the neutron separation energy in the halo system, the $s$-wave effective range  for neutron-core scattering, and a two-body magnetic coupling.  

The electric form factor for $^{11}$Be was calculated previously~\cite{Hammer201117}. The charge radius was found to agree with the known experimental value with the EFT error estimate. We include the magnetic form factor in this analysis. At NLO a two-body magnetic coupling contributes that is fitted to the known magnetic moment. The contribution from the two-body current is consistent with the EFT power counting. We provide the low momentum dependence of the electric and magnetic form factors.

The analysis for $^{19}$C is very similar to $^{11}$Be.  We make a NLO prediction for the charge radius that depends on the effective range determined~\cite{Acharya:2013nia} from $^{19}$C Coulomb dissociation data. For the magnetic form factor we are only able to provide an estimate for the two-body contribution based on the power counting. A determination of the magnetic moment would fix the two-body contribution more precisely.

The power counting for $^{15}$C system is  found to be a little different than the two systems above. We reanalyzed the Coulomb dissociation calculation~\cite{PhysRevC.86.044608}, and found that a non-perturbative $s$-wave  effective range contribution describes that data better. In this system both the effective range and the two-body magnetic coupling contributes at LO. The effective range is determined from a fit to dissociation data, and the two-body current contribution from the magnetic moment. The momentum dependence of the LO electric and magnetic form factors are presented.

\begin{acknowledgments}
The authors thank T. Nakamura for providing data on Coulomb dissociation. We also thank D. R. Phillips for valuable discussions. 
This work is partially supported by the U.S. NSF grant No. PHY-1307453  and  HRD-1345219, and NASA  grant No. NNX09AV07A\end{acknowledgments}


\input{HaloFormFactors_111015.bbl}

\end{document}

%% file: HaloFormFactors_111015.bbl
%